     \documentclass[final,5p,times,twocolumn,authoryear]{elsarticle}
    
    \usepackage{amssymb}
    \usepackage{lipsum}
    \usepackage{graphicx}
    \usepackage{subcaption}
    
    \setcitestyle{square,numbers,sort&compress}
    \usepackage[colorlinks=true,
                linkcolor=blue,
                citecolor=blue,
                urlcolor=blue]{hyperref}
    \usepackage[all]{hypcap}
    \journal{Applied Surface Science}

\begin{document}
    
    \begin{frontmatter}
    
    
    %% \affiliation{organization={},%%            addressline={}, 
    
    \title{Impact of Oxygen Plasma Surface Treatment on Photoresist Adhesion in BaTiO$_3$-Based Photonic Device Fabrication}
    
    
    \author[first]{Weiyou Kong\fnref{equal1}}
    \author[first]{Weijia Kong\fnref{equal1}}
    \author[first]{Lukas Chrostowski}
    \author[first]{Xin Xin\corref{cor1}} 
    \ead{xinx1@ece.ubc.ca}
    \cortext[cor1]{Corresponding author}
    \fntext[equal1]{These authors contributed equally to this work.}
    \affiliation[first]{organization={University of British Columbia},%Department and Organization
                addressline={6200 University Blvd}, 
                city={Vancouver},
                postcode={V6T 1Z4}, 
                state={BC},
                country={Canada}}
    
    \begin{abstract}
    Oxygen-plasma pre-cleans are routine before fabrication, but on BaTiO$_3$ thin films we observed catastrophic photoresist lift-off during mild rinsing/sonication. To explain the failure, we combined optical microscopy, EDS, and XPS. EDS showed no meaningful bulk stoichiometry change, whereas XPS revealed a nanometer-scale, plasma-induced shift in surface chemistry: hydroxylation and carbonate formation consistent with a BaCO$_3$-rich interphase at the resist/BaTiO$_3$ boundary. This chemically weak interphase—recreated upon each plasma step and removable by simple solvent cleaning—provides the mechanism for delamination. The key takeaway for practitioners is process guidance: avoid uncritical O$_2$-plasma use on BTO; if cleaning is required, use alternative chemistries (e.g., UV–ozone) or carefully tuned plasma windows that preserve adhesion. More broadly, the study illustrates how lightweight analytics at the surface (correlative microscopy + surface spectroscopy) can pinpoint the root cause of yield-limiting defects in oxide photonics and translate directly into higher-reliability process recipes.
    \end{abstract}
    
    
    \end{frontmatter}
    
    %\tableofcontents
    
    
    
    \section{Introduction}
    \label{introduction}
    
    Ferroelectric perovskite oxides, such as Barium titanate (BaTiO$_3$, BTO), lithium niobate (LiNbO$_3$), and lithium tantalate (LiTaO$_3$), have attracted significant attention due to their strong electro-optic effect, high dielectric constant, and compatibility with integrated photonic platforms \cite{ref25}\cite{ref38}\cite{ref39}. In recent years, these ferroelectric perovskite oxide thin films have been widely investigated for applications in electro-optic modulators \cite{ref26}, nonlinear optics \cite{ref27}, and tunable photonic devices \cite{ref28}. The integration of them with silicon photonics has further highlighted its potential for low-voltage and high-speed optical signal processing \cite{ref29}\cite{ref30}.
    
    For reliable device fabrication, surface cleanliness plays a critical role. Organic residues, carbon contamination, and adsorbed species can significantly influence the performance and reproducibility of photonic devices \cite{ref18}. Oxygen plasma treatment is a commonly adopted method to remove such surface contaminants, and has been successfully applied to various oxide materials \cite{ref1}. However, plasma exposure is also known to induce surface modifications, including changes in stoichiometry, defect formation, and chemical bonding states \cite{ref2}\cite{ref3}\cite{ref4}\cite{ref10}. These modifications may strongly impact subsequent lithography steps \cite{ref7}\cite{ref20}, yet their effects on BaTiO$_3$ surfaces remain insufficiently explored.
    
    In preliminary tests, we observed that oxygen plasma treatment of BaTiO$_3$ surfaces, although effective in cleaning, unexpectedly resulted in large-scale photoresist delamination during development. To the best of our knowledge, this phenomenon has not been systematically reported in the context of BaTiO$_3$ thin films. Understanding this effect is essential, since photoresist adhesion directly affects the reliability of device patterning and subsequent fabrication steps \cite{ref31}.
    
    In this work, we investigate the influence of oxygen plasma treatment on BaTiO$_3$ thin films with respect to photoresist adhesion and surface chemical modifications. Lithography tests using ZEP520A photoresist were conducted to compare adhesion before and after plasma treatment. X-ray photoelectron spectroscopy (XPS) was employed to identify changes in surface chemical states \cite{ref11}\cite{ref12}\cite{ref17}, while energy-dispersive spectroscopy (EDS) was used to evaluate the bulk composition \cite{ref13}\cite{ref14}\cite{ref15}\cite{ref16}. Our results reveal that oxygen plasma significantly modifies the surface chemistry of BaTiO$_3$, which weakens photoresist adhesion, while leaving the bulk chemical composition unaffected. These findings provide new insights into BaTiO$_3$ surface engineering and highlight the importance of optimizing cleaning processes for reliable device fabrication.
    
    \section{Experimental details}
    \subsection{Specimen preparation}\label{sec:specimen-prep}
    All experiments were performed on the 5 mm $\times$ 5 mm Silicon-on-Insulator (SOI) wafer piece bearing a 300 nm BaTiO$_3$ film (La Luce Cristallina). Surface contaminants were removed by sequential immersion in acetone (ACE, 2 min) and isopropanol (IPA, 2 min) with 50 kHz ultrasonic agitation, followed by nitrogen drying. Oxygen plasma treatment, when applied, was carried out in a Plasma Etch PE-50 etcher at 200 W RF power (O$_2$ flow: 20 sccm, chamber pressure: 300 mTorr, duration: 2 min). For all oxygen-plasma–treated samples, subsequent processing and measurements were initiated within 5 min of completing the plasma step. Unless otherwise specified, all steps were conducted at room temperature using electronic-grade solvents in clean glassware, and nitrogen used for drying was filtered and dry.  
    
    \subsection{Photoresist delamination test}
    ZEP520A positive photoresist was spin-coated at 2000~rpm for 45~s and baked at 180~$^{\circ}$C for 2~min. Film coverage and uniformity were examined by optical microscopy. Adhesion was evaluated by immersing the specimen in deionized water and subjecting it to ultrasonic cleaning (30~s, 50~kHz), as ZEP520A is not water-soluble \cite{ref32} and photoresist loss under these conditions reflects interfacial adhesion failure.  
    
    For the baseline (no-plasma) test, the photoresist was removed by ultrasonication in N-methyl-2-pyrrolidone (NMP, 3~min), followed by ACE (2~min) and IPA (2~min) rinses and nitrogen drying. Immediately after oxygen plasma exposure (parameters described in \hyperref[sec:specimen-prep]{Section~\ref*{sec:specimen-prep}}), ZEP520A was reapplied under identical spin-coating and baking conditions, with minimal exposure to ambient air. The post-plasma specimen was re-examined by optical microscopy, and adhesion was tested under the same deionized-water ultrasonic cleaning (30~s, 50~kHz).  
    
    \subsection{Energy-dispersive spectroscopy (EDS)}
    
    Data were processed and quantified using the Oxford Instruments AZtec software platform (Tru-Q® engine) for automatic peak identification and standardless quantification of Ba, Ti, O, and C. After the baseline measurement, the specimen was exposed to oxygen plasma under the conditions described in \hyperref[sec:specimen-prep]{Section~\ref*{sec:specimen-prep}} and re-measured under identical EDS settings \cite{ref13}.
    
    \subsection{X-ray photoelectron spectroscopy (XPS)}
    Sample preparation and handling followed the unified procedure in \hyperref[sec:specimen-prep]{Section~\ref*{sec:specimen-prep}}. Surface chemical composition and bonding states were analyzed by XPS using a monochromatic Al K$\alpha$ source (hv = 1486.6 eV) \cite{ref11}\cite{ref12}. High-resolution spectra were acquired for the Ba\,3d, Ti\,2p, O\,1s, and C\,1s core levels. Acquisition parameters (step size $\sim$0.05 eV, dwell time $\sim$0.5 s) were selected with a reduced pass energy to ensure high resolution. 
    
    Data were processed in CasaXPS. Background subtraction employed Shirley or Tougaard models as appropriate to the line shape, and component peaks were fitted using Gaussian–Lorentzian line shapes \cite{ref21}. For consistency, the full width at half maximum (FWHM) of the same chemical state was constrained to similar values across different samples, while surface-adsorbate-related components (e.g., –OH, CO$_x$) were allowed moderate variation \cite{ref21}. 
    
    \section{Results}
    \subsection{Photoresist delamination result}
    
    % \hyperref[fig:om_plasma]{Figure~\ref{fig:om_plasma}} presents optical micrographs of a 5\,mm $\times$ 5\,mm BaTiO$_3$ wafer coated with ZEP520A photoresist, comparing conditions without and with oxygen-plasma pretreatment, before and after 30\,s ultrasonication in deionized water. We determine whether photoresist remains by inspecting the uniform central field bounded by the green interference fringe: if this uniform region and its green boundary persist, photoresist is present; if they vanish and only the bare, featureless substrate tone remains, the photoresist has been completely removed.

    % For the wafer without plasma pretreatment (\hyperref[fig:om_plasma:noplasma_before]{Figure~\ref*{fig:om_plasma:noplasma_before}}), only the central area of the small chip was uniformly covered, while the edges showed thicker, non-uniform rings due to chip size effects. After ultrasonication (\hyperref[fig:om_plasma:noplasma_after]{Figure~\ref*{fig:om_plasma:noplasma_after}}), the non-uniform edge rings were largely stripped, yet the central uniform region—together with its green boundary—remained unchanged, indicating sufficient adhesion in the well-coated area.
    
    % In contrast, with oxygen-plasma pretreatment prior to spin-coating (\hyperref[fig:om_plasma:plasma_before]{Figure~\ref*{fig:om_plasma:plasma_before}}), the photoresist adhesion was markedly reduced. Following ultrasonication (\hyperref[fig:om_plasma:plasma_after]{Figure~\ref*{fig:om_plasma:plasma_after}}), the uniform central field and its green boundary disappeared across the chip, leaving only the substrate appearance and confirming complete delamination—even in the previously stable center. This behavior was reproduced on multiple samples, and is summarized quantitatively in \hyperref[tab:resist_retention]{Table~\ref{tab:resist_retention}}, where the retained area of the central uniform region is 100\% for all chips without plasma and 0\% for all chips with plasma.

\begin{figure}[htbp]
      \centering
      \subfloat[\label{fig:om_plasma:noplasma_before}]{
        \includegraphics[width=0.48\linewidth]{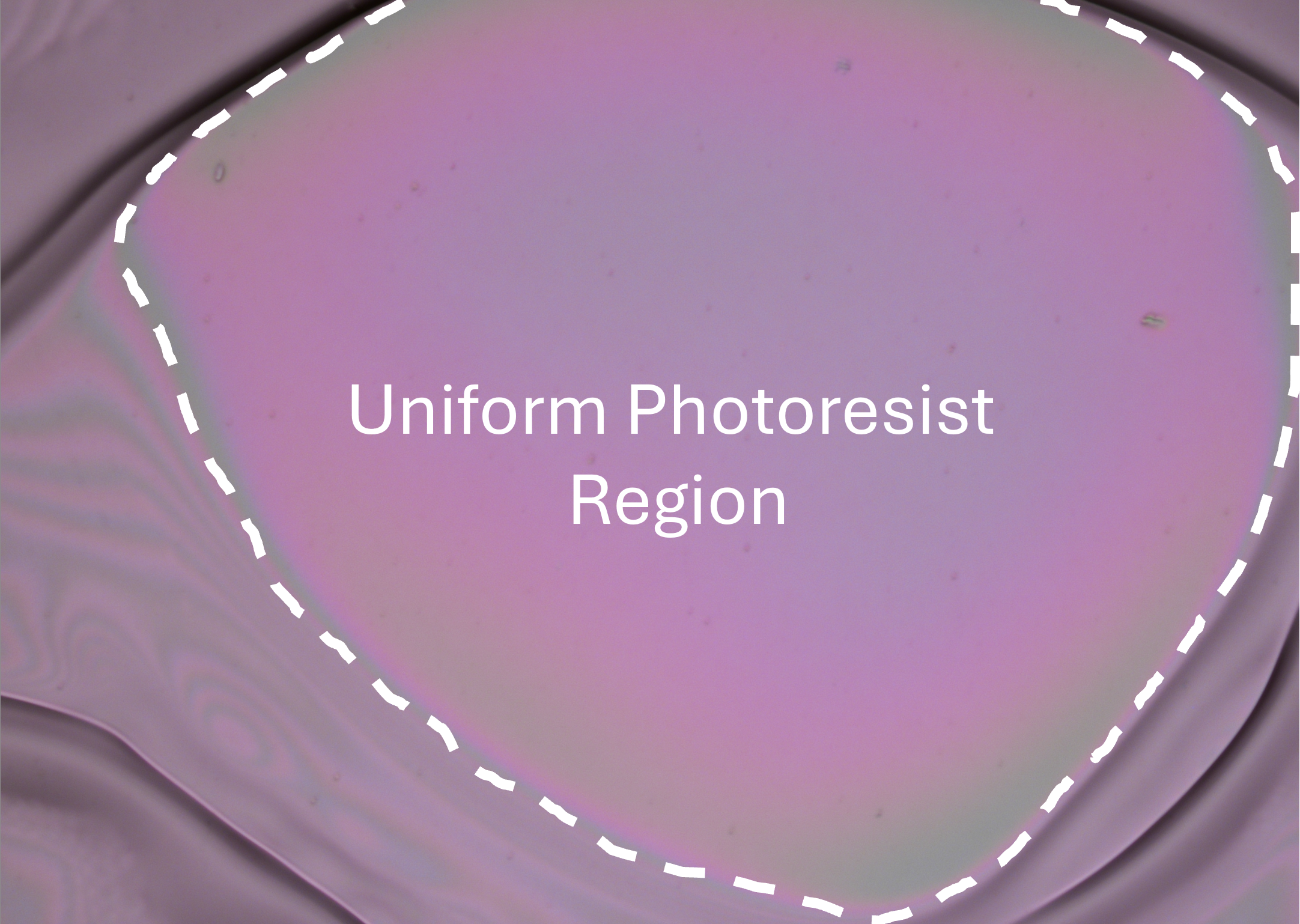}
      }\hfill
      \subfloat[\label{fig:om_plasma:noplasma_after}]{
        \includegraphics[width=0.48\linewidth]{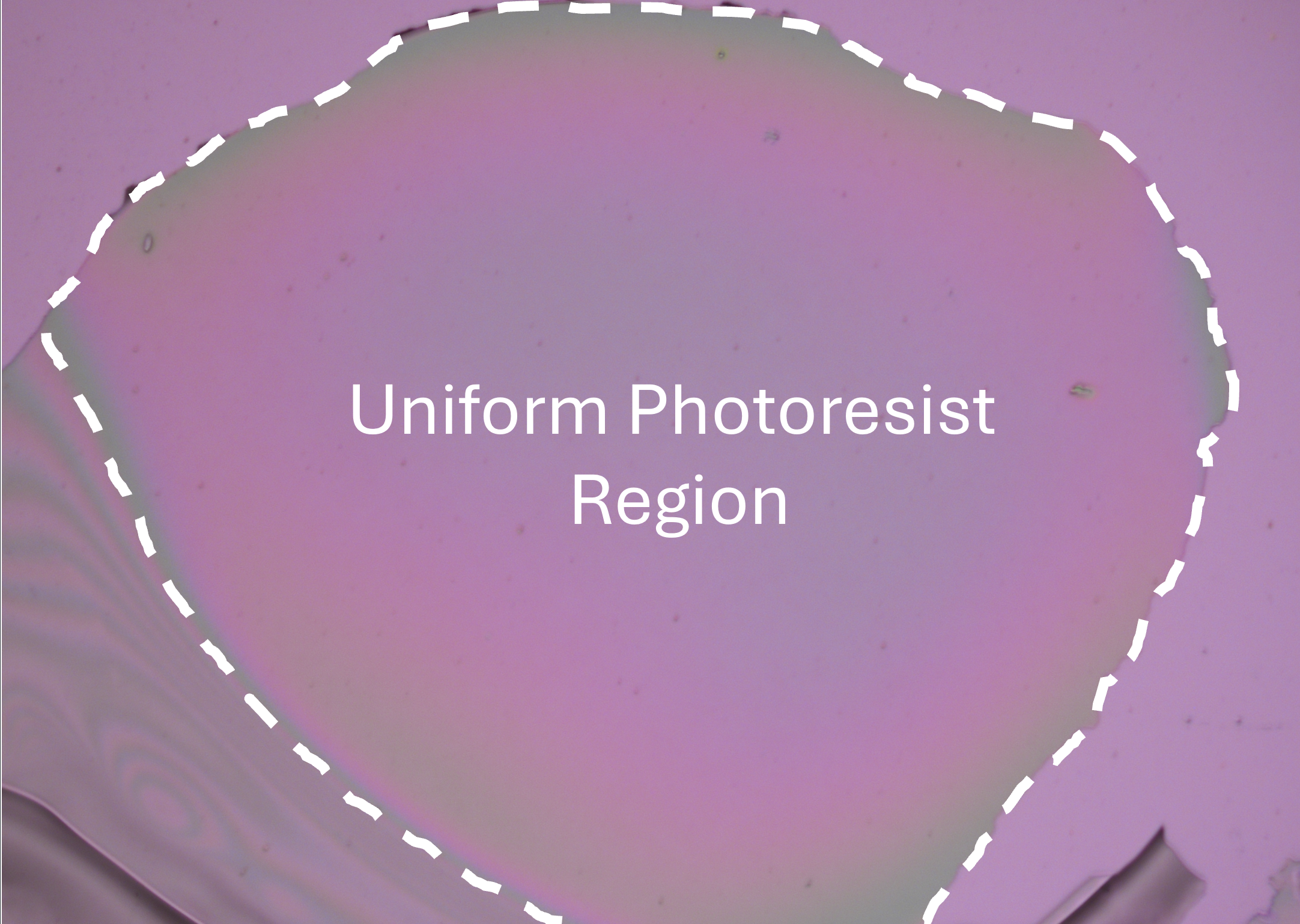}
      }\\[0.6em]
      \subfloat[\label{fig:om_plasma:plasma_before}]{
        \includegraphics[width=0.48\linewidth]{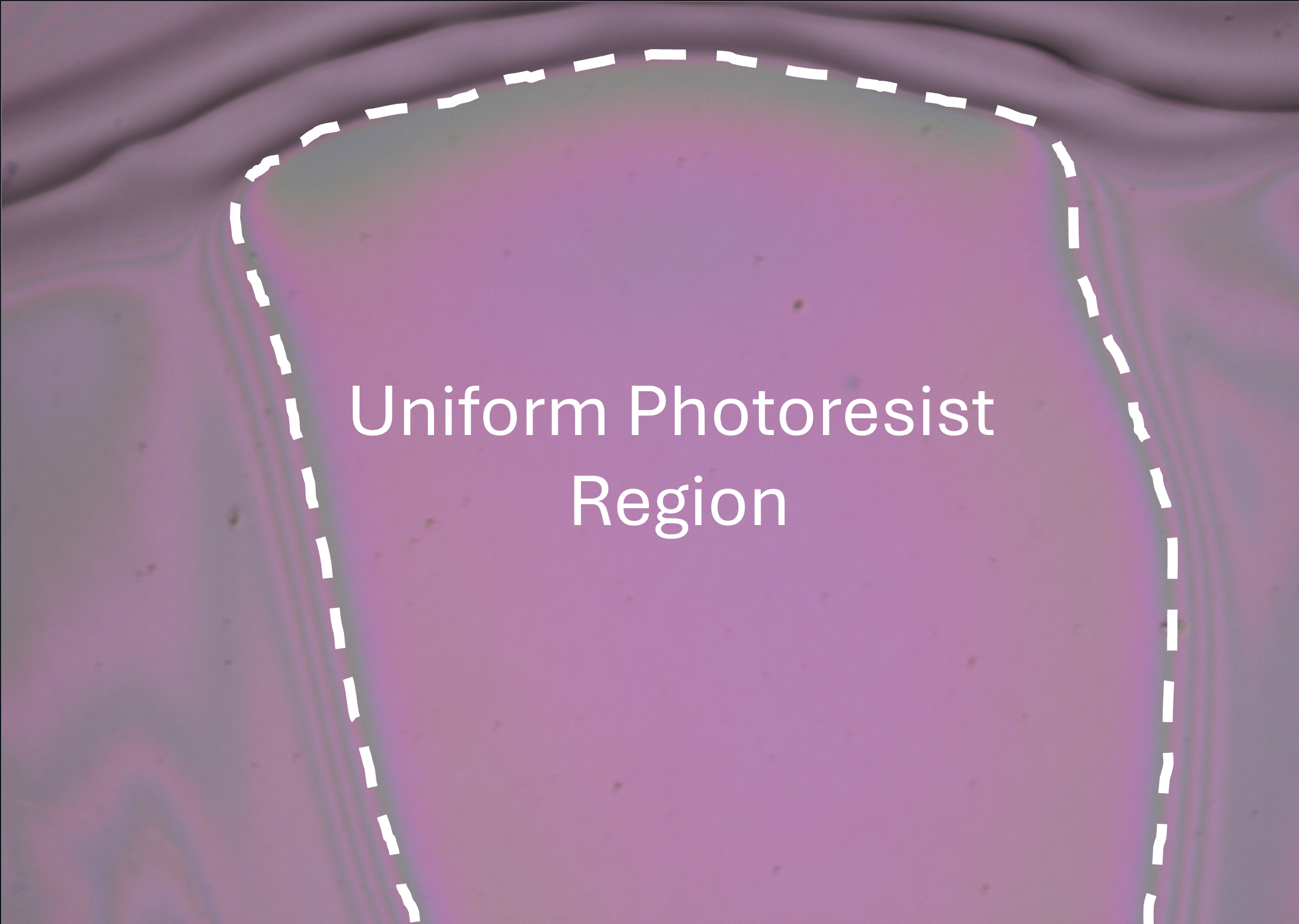}
      }\hfill
      \subfloat[\label{fig:om_plasma:plasma_after}]{
        \includegraphics[width=0.48\linewidth]{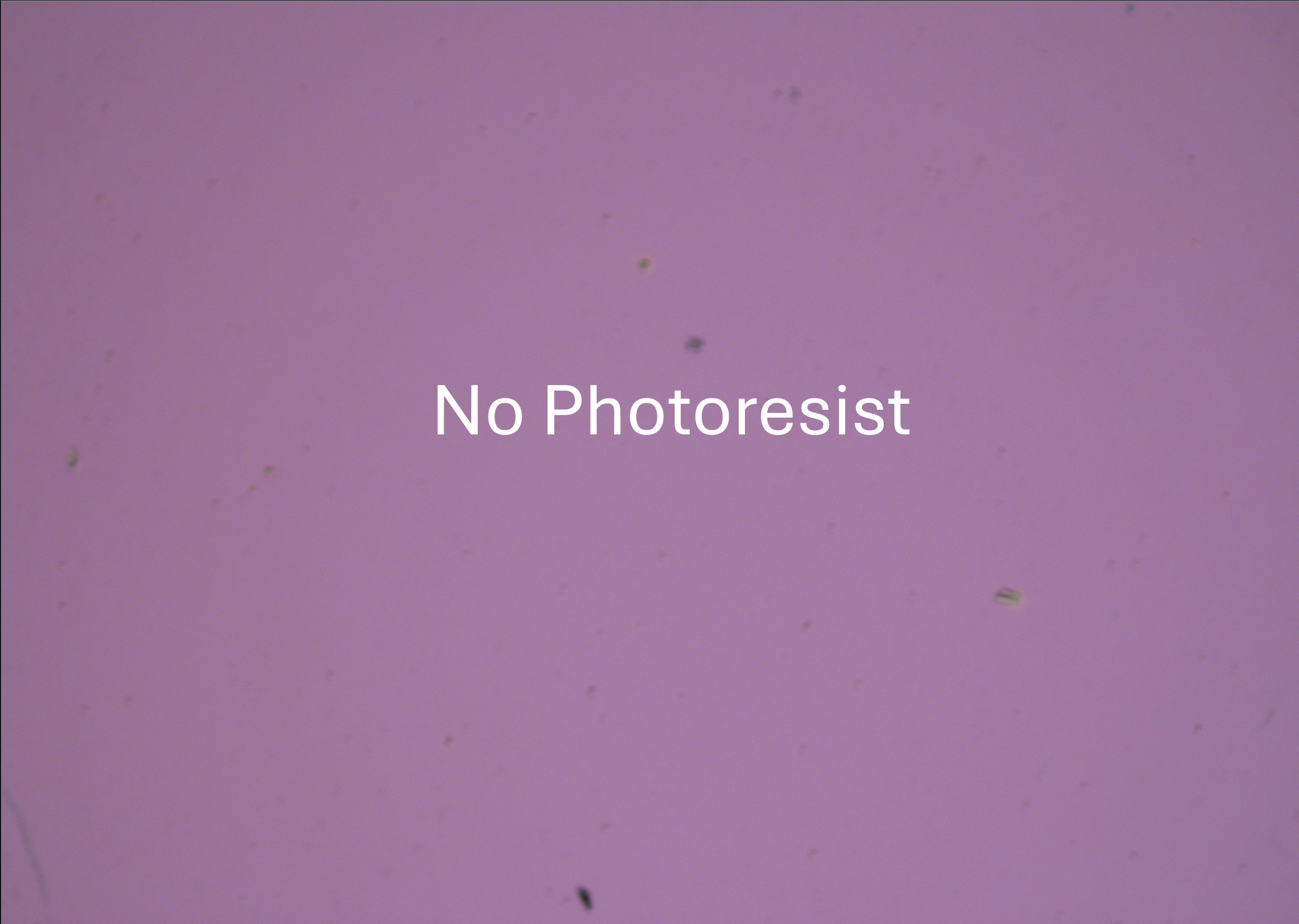}
      }\\[0.6em]
      \caption{Optical micrographs (2.5 mm × 3.5 mm field of view per panel) of a 5 mm × 5 mm BaTiO$_3$ wafer coated with ZEP520A photoresist, comparing conditions without and with oxygen-plasma pretreatment, before and after ultrasonication. Dashed lines outline the Uniform Photoresist Region, whose disappearance indicates complete film removal. (a,b) No plasma: before and after ultrasonication; (c,d) With plasma: before and after ultrasonication. The comparison highlights the loss of photoresist adhesion following oxygen-plasma exposure.}
      \label{fig:om_plasma}
    \end{figure}

    \hyperref[fig:om_plasma]{Figure~\ref{fig:om_plasma}} presents optical micrographs of a 5 mm~$\times$~5 mm BaTiO$_3$ wafer coated with ZEP520A photoresist, comparing samples without and with oxygen-plasma pretreatment, each imaged before and after 30,s ultrasonication in deionized water. We identify the presence of photoresist by the uniform central field bounded by a green interference fringe (outlined in the images): if this uniform region and its green boundary persist, photoresist remains; if both vanish and only the bare, featureless substrate tone is left, the photoresist has been completely removed.

    For the wafer without plasma pretreatment (\hyperref[fig:om_plasma:noplasma_before]{Figure~\ref*{fig:om_plasma:noplasma_before}}), only the central area of the small chip was uniformly coated, while the edges exhibited thicker, non-uniform rings due to chip-size effects. After ultrasonication (\hyperref[fig:om_plasma:noplasma_after]{Figure~\ref*{fig:om_plasma:noplasma_after}}), these non-uniform edge rings were largely stripped, yet the uniform central region—together with its green boundary—remained unchanged, indicating sufficient adhesion where the film was uniform.
    
    In contrast, with oxygen-plasma pretreatment prior to spin-coating (\hyperref[fig:om_plasma:plasma_before]{Figure~\ref*{fig:om_plasma:plasma_before}}), adhesion was markedly reduced. Following ultrasonication (\hyperref[fig:om_plasma:plasma_after]{Figure~\ref*{fig:om_plasma:plasma_after}}), the uniform central field and its green boundary disappeared across the chip, leaving only the substrate appearance and confirming complete delamination—even in the previously stable center. This behavior was reproduced on multiple samples and is summarized quantitatively in \hyperref[tab:resist_retention]{Table~\ref{tab:resist_retention}}, where the retained area of the central uniform region is 100\% for all chips without plasma and 0\% for all chips with plasma.
    
    \begin{table}[h!]
    \centering
    \caption{Retention of the central uniform photoresist region on BaTiO$_3$ chips with and without oxygen-plasma pretreatment, evaluated from optical micrographs after ultrasonication.}
    \label{tab:resist_retention}
    \begin{tabular}{lccc}
    \hline
     & Chip 1 & Chip 2 & Chip 3 \\
    \hline
    Without plasma (\%) & 100 & 100 & 100 \\
    With plasma (\%)    & 0   & 0   & 0   \\
    \hline
    \end{tabular}
    \end{table}
    
    \subsection{EDS result}
    
    \begin{figure}[htbp]
      \centering
      \subfloat[\label{fig:eds_5keV:noplasma}]{
        \includegraphics[width=0.96\linewidth]{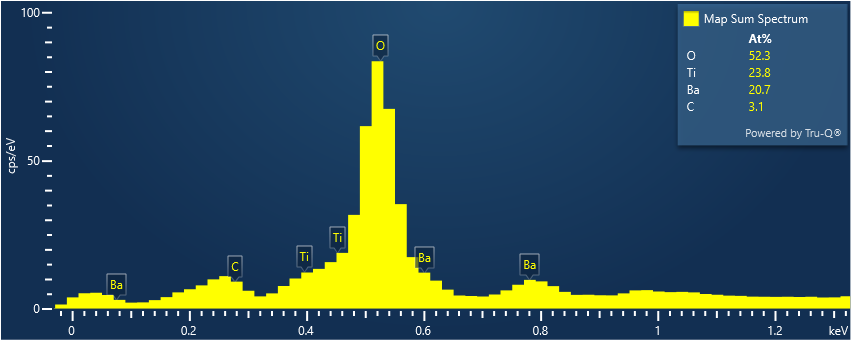}
      }\\
      \subfloat[\label{fig:eds_5keV:plasma}]{
        \includegraphics[width=0.96\linewidth]{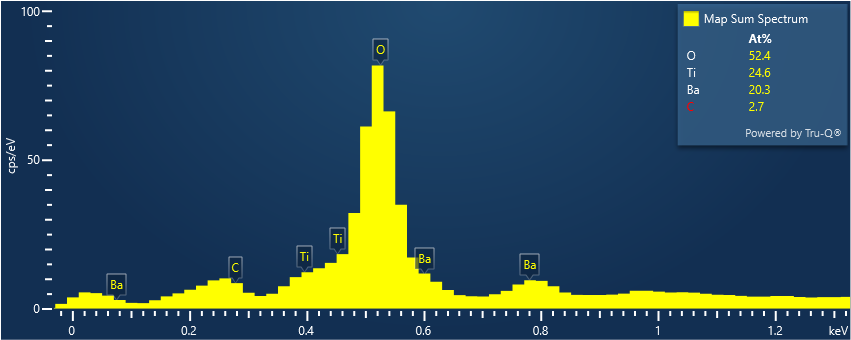}
      }\\[0.6em]
      \caption {EDS mapping of a $5\,{mm} \times 5\,mm$ BaTiO$_3$ sample at an accelerating voltage of $5\,keV$. (a) Spectrum acquired from the as-prepared BaTiO$_3$ surface without oxygen plasma treatment. (b) Spectrum acquired from the BaTiO$_3$ surface after oxygen plasma treatment.}
      \label{fig:eds_5keV}
    \end{figure}
    
    \hyperref[fig:eds_5keV]{Figure~\ref*{fig:eds_5keV}} compares EDS spectra acquired at an accelerating voltage of 5~keV from the same BaTiO$_3$ wafer before and after oxygen–plasma treatment, with panel \hyperref[fig:eds_5keV:noplasma]{Figure~\ref*{fig:eds_5keV:noplasma}} corresponding to the untreated surface and panel \hyperref[fig:eds_5keV:plasma]{Figure~\ref*{fig:eds_5keV:plasma}} to the treated surface. The two spectra display essentially identical peak positions and relative intensities for Ba, Ti, and O, indicating no discernible compositional change within the probed volume. Quantification likewise shows only minor variations within the expected repeatability for oxide EDS: oxygen changes from 52.4~at.\% in the untreated state to 52.3~at.\% after plasma, titanium from 24.6~at.\% to 23.8~at.\%, barium from 20.3~at.\% to 20.7~at.\%, and carbon—reflecting residual adventitious species or background—from 2.7~at.\% to 3.1~at.\%. These small shifts lack a systematic trend and do not constitute a statistically meaningful stoichiometric modification attributable to the plasma under the present acquisition conditions.
    
    \subsection{XPS results}
\begin{figure}[htbp]
      \centering
      \subfloat[Ba 3d\label{fig:xps_noplasma:ba}]{
        \includegraphics[width=0.48\linewidth]{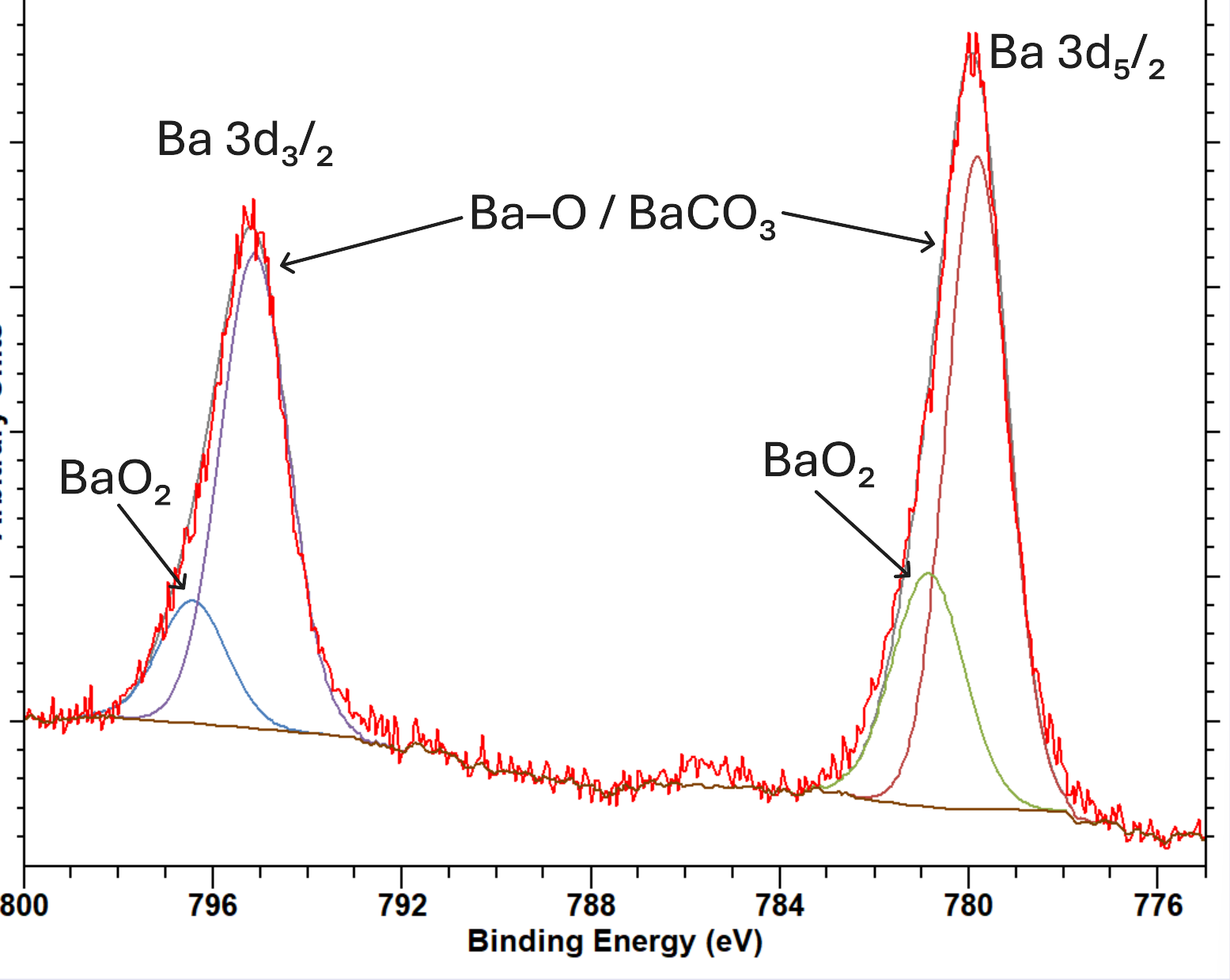}
      }\hfill
      \subfloat[O 1s\label{fig:xps_noplasma:o}]{
        \includegraphics[width=0.48\linewidth]{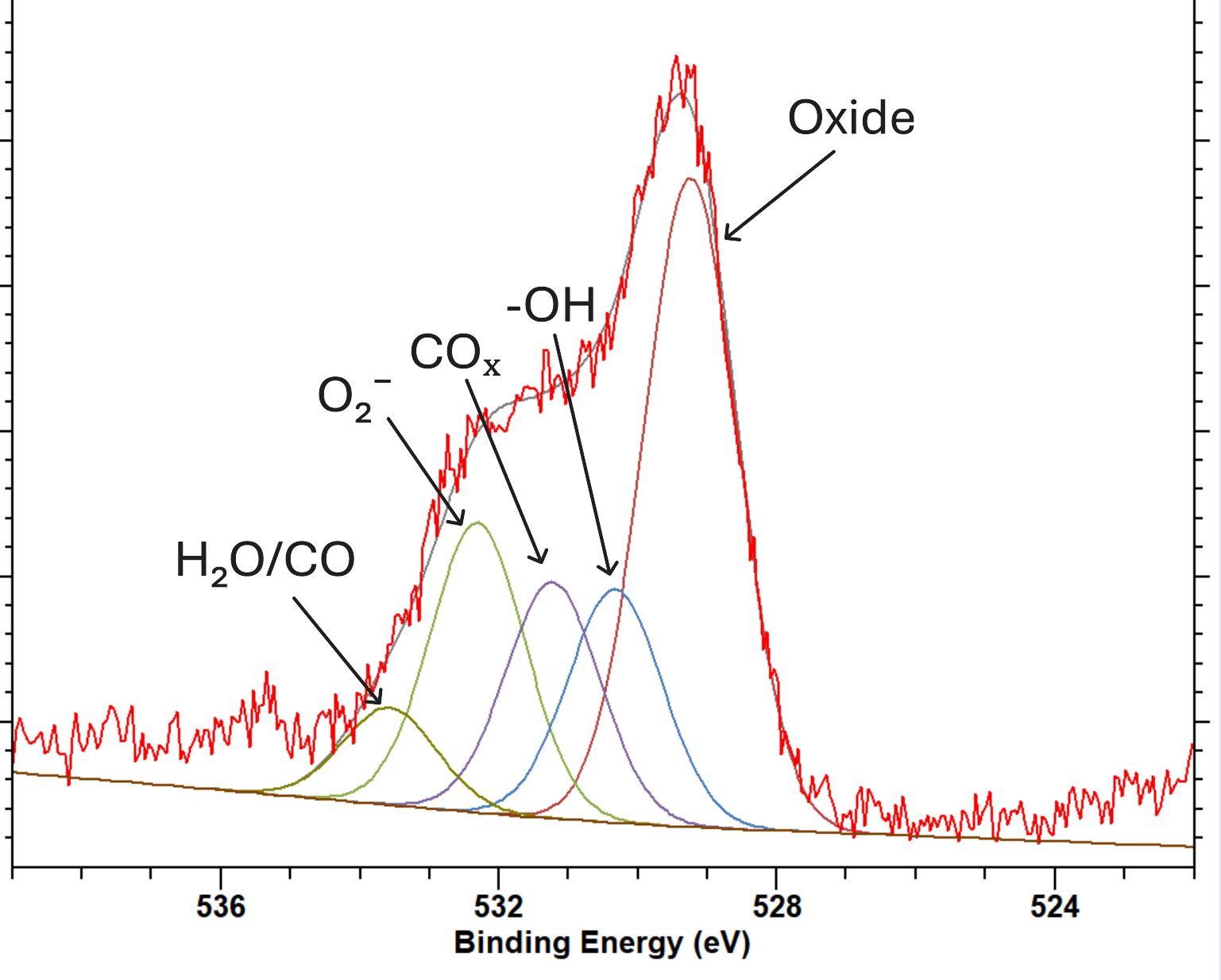}
      }\\[0.1em]
      \subfloat[Ti 2p\label{fig:xps_noplasma:ti}]{
        \includegraphics[width=0.48\linewidth]{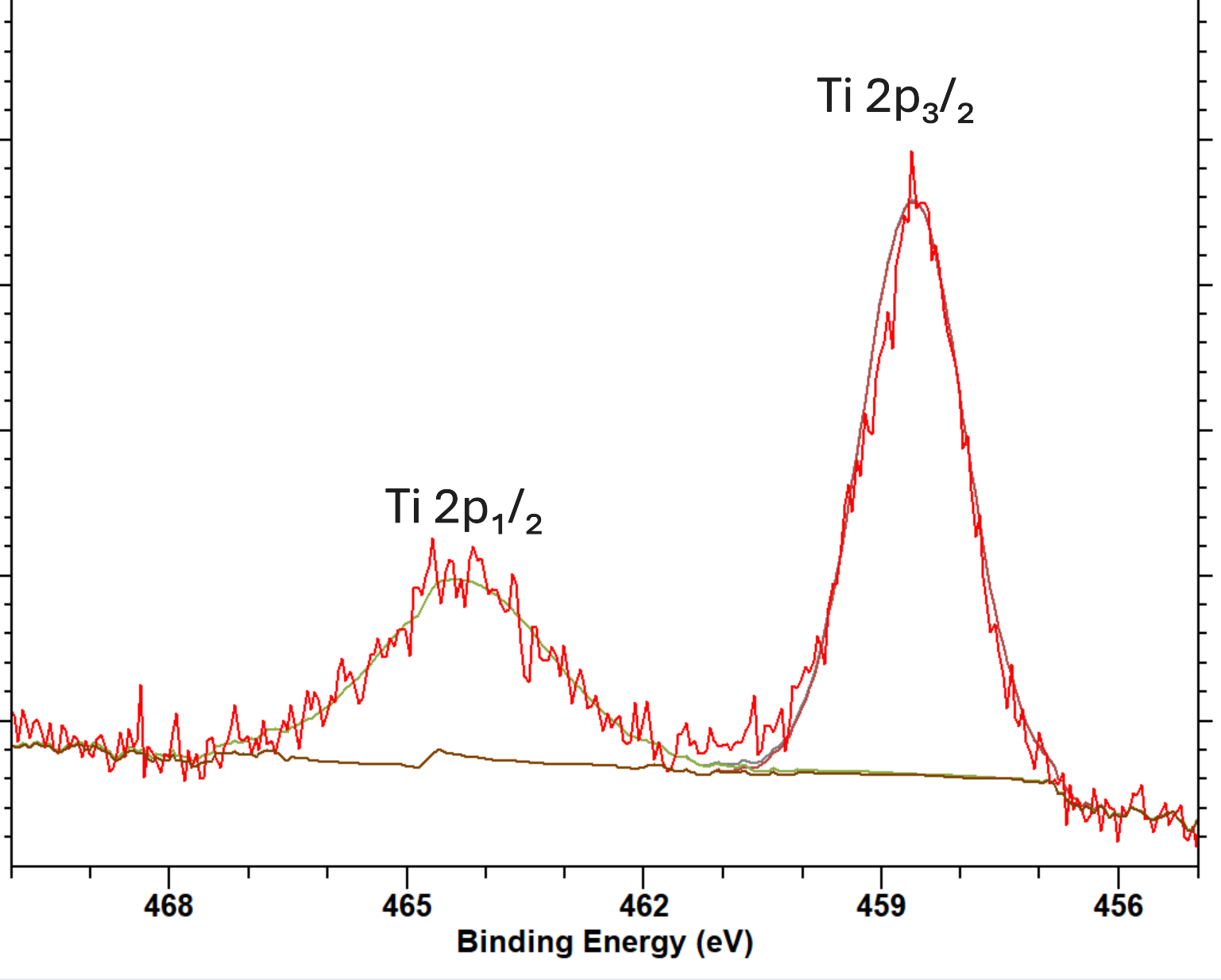}
      }\hfill
      \subfloat[C 1s\label{fig:xps_noplasma:c}]{
        \includegraphics[width=0.48\linewidth]{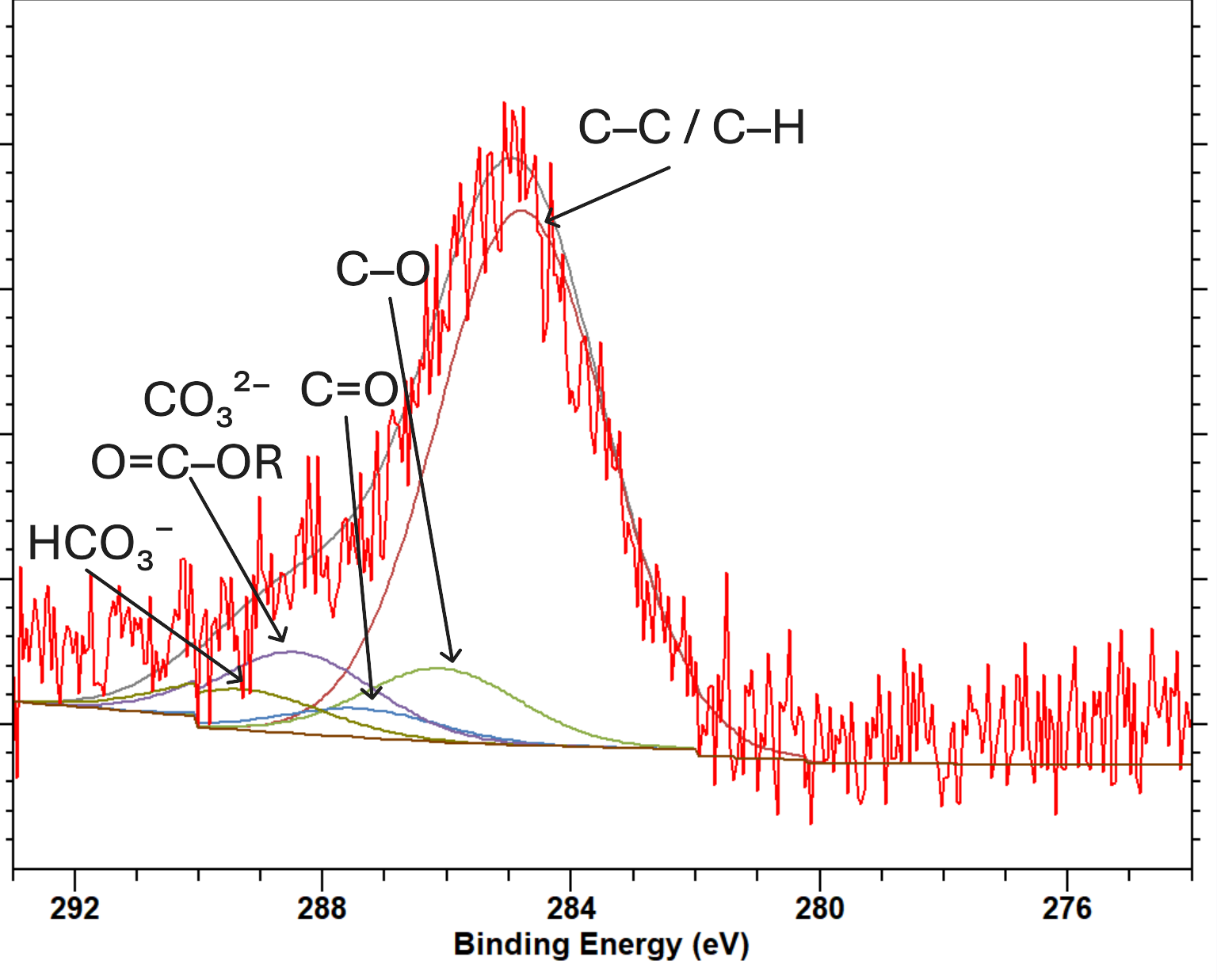}
      }
      \caption{XPS spectra of a BaTiO$_3$ thin film that was sequentially cleaned with propanol and isopropanol and then immediately transferred into the XPS chamber for measurement. Panels: (a) Ba\,3d, (b) O\,1s, (c) Ti\,2p, (d) C\,1s.}
        \label{fig:xps_noplasma}
    \end{figure}

    The acquired spectra are summarized in \hyperref[fig:xps_noplasma]{Figure~\ref*{fig:xps_noplasma}}, where the main regions of interest include Ba\,3d, O\,1s, Ti\,2p, and C\,1s. For the Ba\,3d region \hyperref[fig:xps_noplasma:ba]{Figure~\ref*{fig:xps_noplasma:ba}}, the characteristic doublet 3d$_{5/2}$ and 3d$_{3/2}$ is clearly identified, positioned near 780\,eV and 795\,eV with a separation of about 15\,eV. The lower-binding-energy part of the profile consists of a main contribution from BaO and BaCO$_3$, which cannot be fully separated at the present resolution \cite{ref2}. In addition, a weaker component shifted by roughly +1.3 to +1.5\,eV is observed, attributed to BaO$_2$. The higher-energy 3d$_{3/2}$ peak follows the same two-component structure, shifted upwards by the spin–orbit splitting \cite{ref22}\cite{ref23}.
    
    Turning to the O\,1s spectrum \hyperref[fig:xps_noplasma:o]{Figure~\ref*{fig:xps_noplasma:o}}, the strongest signal at about $529\,\mathrm{eV}$ corresponds to the oxide. Several additional features arise at higher binding energies. A shoulder about +1.3\,eV higher originates from hydroxyl groups on the surface, including both terminal and bridging OH species. The next contribution at $\Delta\mathrm{BE}\approx+2.1\,\mathrm{eV}$ is linked to carbonate and other carbon–oxygen environments such as esters or carboxyl groups \cite{ref5}\cite{ref18}. A distinct feature at $\Delta\mathrm{BE}\approx+3.2\,\mathrm{eV}$ is related to chemisorbed oxygen species, often described as peroxo or superoxo complexes, and bears similarity to the oxygen found in BaO$_2$. Finally, the broad high-energy tail around $\Delta\mathrm{BE}\approx+4.5\,\mathrm{eV}$ is assigned to adsorbed molecular water, with possible overlap from C--O groups \cite{ref2}.
    
    In the Ti\,2p spectrum \hyperref[fig:xps_noplasma:ti]{Figure~\ref*{fig:xps_noplasma:ti}}, the spin–orbit pair is resolved as expected. The 2p$_{3/2}$ peak centers at approximately 458.7\,eV, indicating Ti in the $+4$ oxidation state \cite{ref23}. If oxygen vacancies were present, a shoulder at lower binding energy corresponding to Ti$^{3+}$ would be expected. However, such a feature is not discernible in our data, suggesting no significant Ti$^{3+}$ contribution within the detection limit \cite{ref2}.
    
    Finally, in the C\,1s region \hyperref[fig:xps_noplasma:c]{Figure~\ref*{fig:xps_noplasma:c}}, four distinct components can be resolved. The main peak at about 284.7\,eV arises from C--C and C--H bonds. At slightly higher binding energies, features at $\Delta\mathrm{BE}\approx+1.5$ and $+2.6\,\mathrm{eV}$ indicate carbon species bonded to oxygen through either C--O or C{=}O linkages. A more oxidized contribution is detected at $\Delta\mathrm{BE}\approx+3.8\,\mathrm{eV}$, attributed to carbonate, ester, or carboxylate groups \cite{ref18}. At the highest energies, a weaker signal at $\Delta\mathrm{BE}\approx+4.65\,\mathrm{eV}$ can be associated with bicarbonate species, though this component is less prominent.
    
\begin{figure}[htbp]
      \centering
      \subfloat[Ba 3d\label{fig:xps_plasma:ba}]{
        \includegraphics[width=0.48\linewidth]{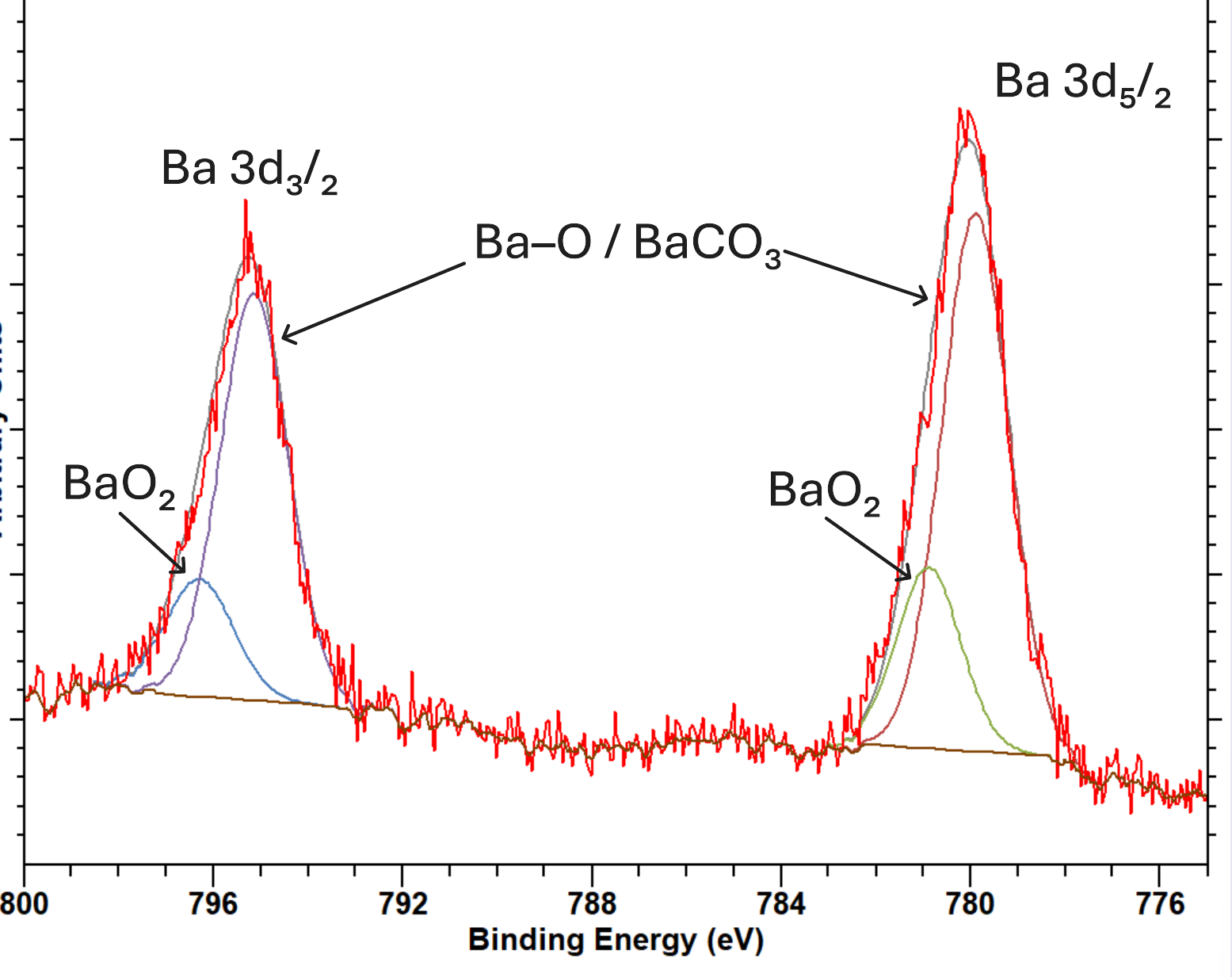}
      }\hfill
      \subfloat[O 1s\label{fig:xps_plasma:o}]{
        \includegraphics[width=0.48\linewidth]{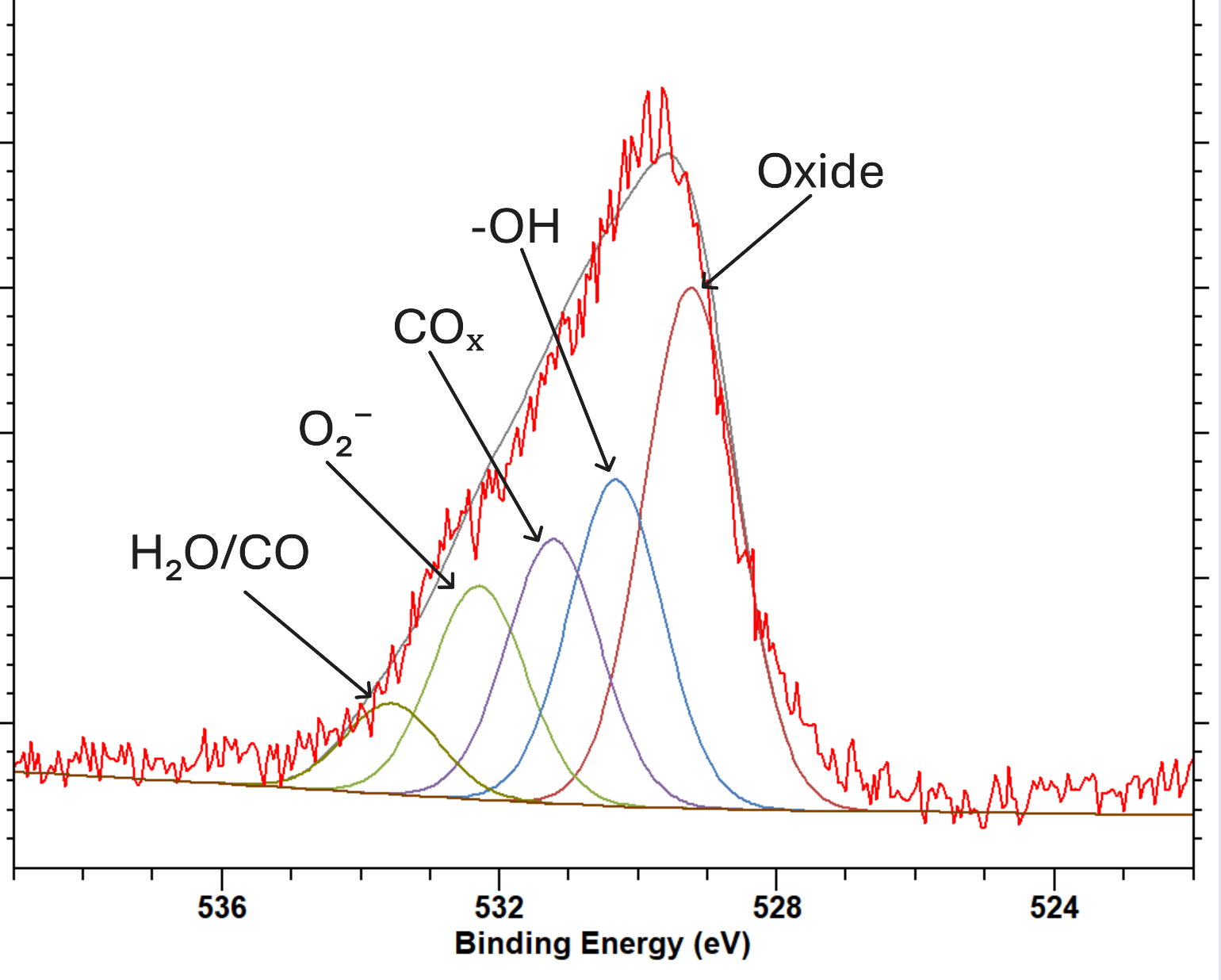}
      }\\[0.6em]
      \subfloat[Ti 2p\label{fig:xps_plasma:ti}]{
        \includegraphics[width=0.48\linewidth]{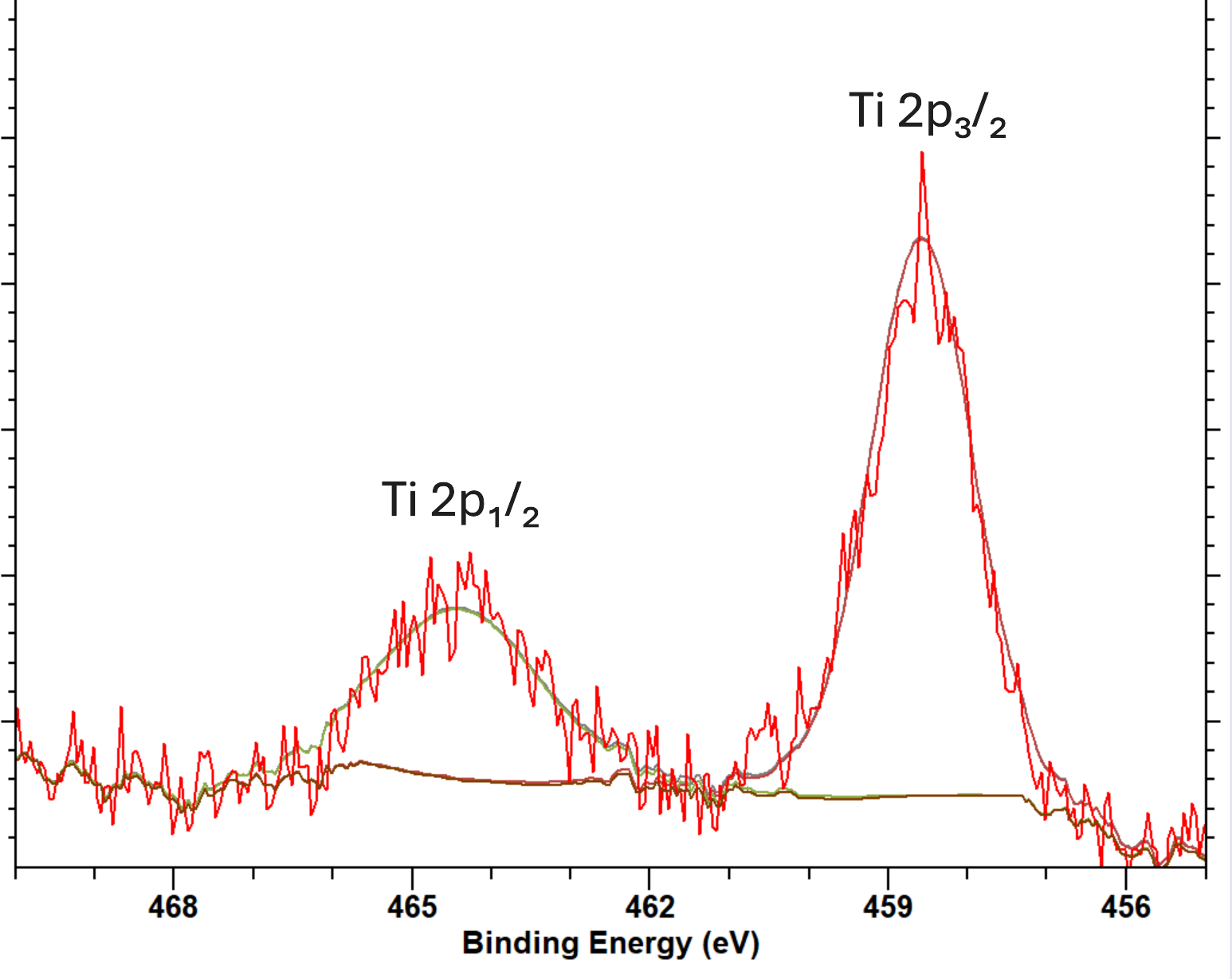}
      }\hfill
      \subfloat[C 1s\label{fig:xps_plasma:c}]{
        \includegraphics[width=0.48\linewidth]{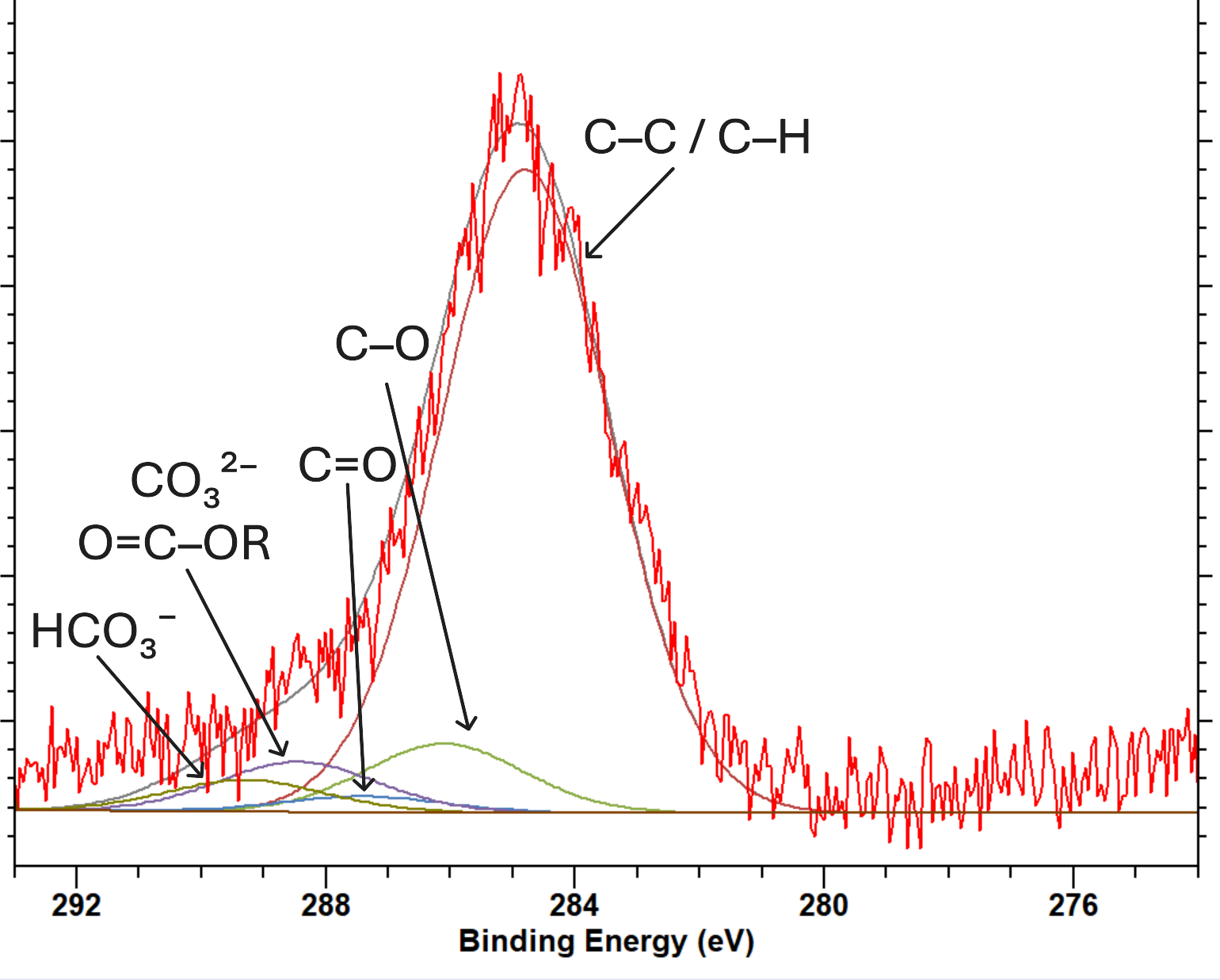}
      }
      \caption{XPS spectra of a BaTiO$_3$ thin film after cleaning with propanol and isopropanol, followed by an additional O$_2$ plasma surface treatment before immediate transfer into the XPS chamber. Panels: (a) Ba\,3d, (b) O\,1s, (c) Ti\,2p, (d) C\,1s.}
      \label{fig:xps_plasma}
    \end{figure}

    After completing the initial measurements, the BaTiO$_3$ wafer was subjected to the same sequential cleaning procedure using ACE and IPA. In this case, however, an additional step was introduced: the sample was treated with oxygen plasma prior to the XPS measurement. Immediately after the plasma treatment, the wafer was transferred into the XPS chamber as quickly as possible to minimize air exposure, and the same set of scans was carried out. The corresponding spectra are shown in \hyperref[fig:xps_plasma]{Figure~\ref*{fig:xps_plasma}}, where the regions Ba\,3d, O\,1s, Ti\,2p, and C\,1s were again analyzed.
    
    By comparing the four regions in \hyperref[fig:xps_noplasma]{Figure~\ref*{fig:xps_noplasma}} and \hyperref[fig:xps_plasma]{Figure~\ref*{fig:xps_plasma}}, it is evident that the Ba\,3d, Ti\,2p, and C\,1s spectra do not show any major differences relative to the no-plasma case. In contrast, the O\,1s region exhibits a clear change: the relative intensities of the CO$_x$ and OH components increase significantly. Before plasma treatment, the O$_2^-$ species represented the largest fraction of the surface-related oxygen signals aside from the oxide. After plasma exposure, however, the CO$_x$ and OH contributions both surpass O$_2^-$ in intensity \cite{ref18}\cite{ref24}. This observation indicates that oxygen plasma processing generates new chemical species on the BaTiO$_3$ surface, which is consistent with the observed detachment of photoresist from the treated regions.

    \section{Discussion}
    \subsection{Photoresist delamination and its reversibility}
    Optical observations demonstrate that oxygen-plasma pretreatment severely deteriorates the adhesion of ZEP520A on BaTiO$_3$. After plasma exposure, the photoresist detaches even in the previously stable central region during mild ultrasonication.

    It is noteworthy that if the plasma-treated wafer was stripped of photoresist and subsequently re-cleaned with ACE and IPA, the photoresist could once again be spin-coated without delamination. However, repeating the oxygen-plasma pretreatment restored the delamination issue.
    
    Taken together, these observations demonstrate that oxygen-plasma pretreatment induces persistent surface modifications on BaTiO$_3$ that strongly degrade photoresist adhesion. The delamination is not attributable to a transient contamination effect, but rather to stable chemical changes at the BaTiO$_3$ surface that reproducibly weaken the photoresist–substrate interface.

    \subsection{EDS evidence: absence of bulk compositional change}
    EDS at 5~keV shows nearly identical Ba, Ti, and O peak intensities and atomic ratios before and after plasma treatment.

    The absence of a measurable change is consistent with the depth sensitivity of EDS at 5~keV. In dense oxides, the characteristic X-ray signal is generated primarily from a near-surface interaction volume extending to a depth of approximately 50–100~nm \cite{ref33}. Given that 2 minutes oxygen plasma treatment is expected to modify only the outermost few nanometers to a few tens of nanometers, any plasma-induced alteration would be strongly diluted by the much larger contribution from the underlying bulk within the EDS sampling depth \cite{ref34}. Therefore, the near-invariance of the 5~keV EDS results supports the interpretation that any plasma-related modifications, if present, are confined to a substantially thinner surface region than the EDS information depth, rather than indicating a bulk compositional change.
    
    To probe such ultrathin surface modifications directly, we complement EDS with XPS, whose intrinsic surface sensitivity is better suited to detect nanometer-scale changes in chemical states and bonding at the extreme surface of BaTiO$_3$ \cite{ref11}.
    
    \subsection{XPS analysis: surface chemistry and carbonate formation}
    XPS, which is intrinsically surface sensitive, reveals substantial increases in hydroxyl and carbonate species following plasma exposure. The appearance of additional O~1s components at higher binding energies and shifts in Ba~3d peaks indicate the formation of BaO-rich terminations.
    
    To further investigate this effect, the same sample was again cleaned with propanol and isopropanol and subjected to the same XPS procedure. The resulting spectra closely resemble those of the no-plasma reference, confirming that the newly formed surface species can be effectively removed by solvent cleaning. To ensure reproducibility and exclude sample-to-sample variations, two additional BaTiO$_3$ wafers were processed under identical conditions, and in both cases the results were essentially identical to those described above.
    
    Quantitative deconvolution of the O\,1s region provides additional insight into the relative growth of hydroxyl and carbonate-related components. As summarized in \hyperref[tab:xps_growth]{Table~\ref*{tab:xps_growth}}, the --OH signal increases by approximately 73.49\%, 51.43\%, and 37.23\% for Chips~1--3, respectively, while the CO$_x$ contribution rises by 36.53\%, 42.86\%, and 17.73\%. The consistent upward trend across all samples corroborates that oxygen plasma treatment enhances both hydroxylation and surface carbonate formation, with chip-to-chip variations plausibly arising from local stoichiometry, surface roughness, or initial termination differences.
    
    \begin{table}[h!]
    \centering
    \caption{Relative growth rates of hydroxyl and carbonate species derived from O~1s XPS peak deconvolution after plasma treatment.}
    \label{tab:xps_growth}
    \begin{tabular}{lccc}
    \hline
     & \textbf{Chip 1} & \textbf{Chip 2} & \textbf{Chip 3} \\
    \hline
    Growth Rate: --OH & 73.49\% & 51.43\% & 37.23\% \\
    Growth Rate: CO$_x$ & 36.53\% & 42.86\% & 17.73\% \\
    \hline
    \end{tabular}
    \end{table}
    
    This increase can be rationalized by considering the surface chemistry induced by oxygen plasma. The energetic oxygen species generated during plasma exposure partially disrupt the Ba--Ti--O framework at the surface, giving rise to BaO-rich terminations. Once formed, BaO is highly reactive and, even under brief ambient exposure during sample transfer, undergoes rapid reactions with atmospheric components \cite{ref3}\cite{ref5}. In particular,
    \[
    \mathrm{BaO} + \mathrm{H_2O} \;\to\; \mathrm{Ba(OH)_2}, \qquad
    \mathrm{BaO} + \mathrm{CO_2} \;\to\; \mathrm{BaCO_3}.
    \]
    
    These reactions yield hydroxyl and carbonate species that manifest as enhanced OH and CO$_x$ contributions in the O\,1s region after plasma treatment.

    \subsection{Mechanism: plasma-induced \texorpdfstring{BaCO\(_3\)}{BaCO3} interlayer causes resist delamination}
    \label{subsec:plasma_baco3_mechanism}
    
    O$_2$ plasma modifies only the outermost BaTiO$_3$ surface; subsequent air exposure yields a nanometre-scale carbonate termination (predominantly BaCO$_3$) located between the photoresist and the substrate \cite{ref18}\cite{ref3}\cite{ref24}\cite{ref6}. This interphase is confined to the depth range probed by XPS and remains below the typical information depth of 5\,keV EDS, consistent with an interfacial---rather than bulk---origin of the failure.
    
    Hydroxyl termination [Ba(OH)$_2$ or surface –OH] is generally associated with increased surface polarity and improved wettability (cf.\ Si–OH on plasma-treated Si) \cite{ref35} and, by itself, is not expected to degrade adhesion to polar photoresists. The delamination is therefore attributed to the carbonate interphase, which exhibits poor chemical compatibility with the organic photoresist and acts as a low-coupling "slip" layer along which cracks nucleate and propagate during baking, rinsing, or mild ultrasonication.
    
    No claim is made regarding an intrinsic mechanical or chemical ``weakness'' of BaCO$_3$. The description of the interphase as \emph{process-fragile} follows from two observations: (i) interfacial lift-off appears only after plasma treatment, and (ii) brief ACE/IPA rinses remove the carbonate termination and restore stable coating. Collectively, these findings identify a plasma-induced BaCO$_3$ interphase as the proximate cause of photoresist delamination, whereas Ba(OH)$_2$ termination alone is unlikely to be responsible.

    \subsection{Implications for ferroelectric oxide photonic fabrication}
    The present study reveals that oxygen plasma, although widely used in standard CMOS fabrication for surface cleaning \cite{ref36}, can become unreliable when applied to ferroelectric oxides such as BaTiO$_3$. The formation of a thin BaCO$_3$-rich surface layer after plasma exposure leads to a complete loss of photoresist adhesion, indicating that BaTiO$_3$ is not fully compatible with conventional O$_2$-plasma cleaning processes.
    
    Beyond BaTiO$_3$, these results serve as an important reference for other ferroelectric oxides such as LiNbO$_3$ or LiTaO$_3$. While their detailed reactions may differ, similar unexpected effects could arise when they are exposed to plasma environments. The BaTiO$_3$ case thus provides a clear example of how plasma-induced chemistry can undermine fabrication reliability, highlighting the need to explore alternative or modified cleaning approaches—such as UV–ozone \cite{ref37} treatments—better suited for ferroelectric oxide photonic device manufacturing.

    \subsection{Future work}
    \label{subsec:future_work}
    
    While this study establishes a clear link between O$_2$ plasma pretreatment and photoresist delamination on BaTiO$_3$ thin films, several avenues remain open for further investigation. Future experiments will explore whether the observed adhesion loss is specific to ZEP520A or generalizable to other classes of photoresists. Testing novel photoresists with different chemical backbones and polarity could provide insight into photoresist–substrate compatibility and potentially identify formulations more tolerant to BaTiO$_3$ surface modifications.

    Beyond treatment duration, adjusting the RF power applied during O$_2$ plasma exposure will help determine whether there exists a threshold or safe operating window that enables surface cleaning without triggering adhesion failure. In parallel, the effect of plasma exposure duration will be systematically varied to probe the threshold conditions under which adhesion failure emerges. Such studies can establish processing windows that minimize deleterious surface chemistry while retaining cleaning effectiveness.
    
    Additionally, distinguishing whether the interfacial layer responsible for adhesion loss is dominated by BaCO$_3$ or Ba(OH)$_2$ formation remains an important open question. Future work employing depth-resolved spectroscopy (e.g., angle-resolved XPS, ToF-SIMS, or in situ infrared spectroscopy) and controlled post-plasma atmospheres (CO$_2$-free vs. H$_2$O-rich) will be necessary to isolate the chemical pathway that most directly weakens resist adhesion.
    
    These future investigations will not only deepen the mechanistic understanding of plasma–BaTiO$_3$ interactions but also enable the development of optimized surface-preparation protocols tailored to reliable lithography in BaTiO$_3$-based integrated photonics

    \section{Summary and conclusions}
    In this work, we systematically investigated the influence of oxygen plasma treatment on BaTiO$_3$ thin films prior to lithographic processing. Although plasma pretreatment is widely employed for surface cleaning, our experiments revealed that it induces severe photoresist delamination. Optical microscopy and ultrasonic testing confirmed that photoresist adhesion was strongly degraded after plasma exposure, in contrast to stable adhesion on untreated surfaces.  
    
    Complementary spectroscopic analyses provided further insight. EDS showed no discernible change in bulk stoichiometry. XPS, however, revealed significant increases in hydroxyl and carbonate species following plasma treatment,  indicating that oxygen plasma modifies only the extreme surface. These findings demonstrate that oxygen plasma generates a BaCO$_3$-rich interphase confined to the outermost nanometers of the BaTiO$_3$ surface.
    
    The novelty of this study lies in the first systematic demonstration that how O$_2$ plasma pretreatment affects BaTiO$_3$ thin film, despite its widespread use for surface cleaning, directly causing large-area photoresist delamination on BaTiO$_3$ thin films. Unlike prior reports that focused solely on the presence of carbonate species, we establish the causal link between plasma-induced surface chemistry and practical lithographic failure, and further show that a simple ACE/IPA rinse can effectively restore adhesion.  
    
    From an engineering standpoint, these findings provide actionable guidance for the fabrication of BaTiO$_3$-based photonic devices. They highlight the risk of adopting standard oxide-cleaning protocols without adaptation, and they suggest practical mitigation strategies that enable reliable photoresist coating and pattern transfer. By identifying a previously overlooked process fragility, this work offers both mechanistic understanding and immediate value for improving yield and reproducibility in ferroelectric integrated photonics.

    \section*{Acknowledgements}
    This work was supported by the Natural Sciences and Engineering Research Council of Canada (NSERC) Ferroelectric Consortium. We gratefully acknowledge the UBC Berlinguette Group for providing training and access to the XPS facility. 
    We also thank the UBC AMPEL facility for access to EDS equipment and cleanroom resources. 
    Special thanks are extended to Robin Kim and Dr. Vien Van for their valuable advice and insightful suggestions.

    \section*{AI Assistance Statement}
    During the preparation of this work the authors used ChatGPT in order to improve the clarity, conciseness, and language fluency of the manuscript. After using this tool, the authors reviewed and edited the content as needed and take full responsibility for the content of the publication.

    \bibliographystyle{elsarticle-num} 
    \bibliography{example}
    
    
    
    %% For example:
    
    %% \bibitem[Aladro et al.(2015)]{Aladro15} Aladro, R., Martín, S., Riquelme, D., et al. 2015, \aas, 579, A101

    %%\end{thebibliography}
    
    \end{document}